\documentclass[journal=jacsat,manuscript=article]{achemso}

\usepackage{setspace}
\usepackage{chemformula} % Formula subscripts using \ch{}
\usepackage[T1]{fontenc} % Use modern font encodings
\usepackage{caption}
\usepackage{subcaption}
\usepackage{cleveref}
\usepackage{soul}
\setstcolor{red}
\author{Nasrin Eyvazi}
\affiliation{Department of Physics, Institute for Advanced Studies in Basic Sciences (IASBS), Zanjan 45137-66731, Iran}
\author{Davood Abbaszadeh}
\affiliation{Department of Physics, Institute for Advanced Studies in Basic Sciences (IASBS), Zanjan 45137-66731, Iran}
\author{Morad Biagooi}
\affiliation{Intelligent Data Aim Ltd (IDA Ltd), Science and Technology Park of Institute for Advanced studies in Basic Sciences, Zanjan 45137-65697, Iran}
\author{SeyedEhsan Nedaaee Oskoee}
\affiliation{Department of Physics, Institute for Advanced Studies in Basic Sciences (IASBS), Zanjan 45137-66731, Iran}
\alsoaffiliation{Research Center for Basic Sciences \& Modern Technologies (RBST), Institute for Advanced Studies in Basic Sciences (IASBS), Zanjan 45137-66731, Iran}
\email{nedaaee@iasbs.ac.ir}
\phone{(+98) 241-415-2217}
\fax{(+98) 241-415-2104}

\title[An \textsf{achemso} demo]
  {Performance investigation of supercapacitors with PEO-based gel polymer \& ionic liquid electrolytes: Molecular Dynamics Simulation}

\abbreviations{IR,NMR,UV}
\keywords{American Chemical Society, \LaTeX}

\begin{document} 
	%\setstretch{3.5}
	
%%%%%%%%%%%%%%%%%%%%%%%%%%%%%%%%%%%%%%%%%%%%%%%%%%%%%%%%%%%%%%%%%%%%%
%% The "tocentry" environment can be used to create an entry for the
%% graphical table of contents. It is given here as some journals
%% require that it is printed as part of the abstract page. It will
%% be automatically moved as appropriate.
%%%%%%%%%%%%%%%%%%%%%%%%%%%%%%%%%%%%%%%%%%%%%%%%%%%%%%%%%%%%%%%%%%%%%
%%%%%%%%%%%%%%%%%%%%%%%%%%%%%%%%%%%%%%%%%%%%%%%%%%%%%%%%%%%%%%%%%%%%%
%% The abstract environment will automatically gobble the contents
%% if an abstract is not used by the target journal.
%%%%%%%%%%%%%%%%%%%%%%%%%%%%%%%%%%%%%%%%%%%%%%%%%%%%%%%%%%%%%%%%%%%%%
\begin{abstract}
Due to the importance of using supercapacitors in electronic storage devices, improving their efficiency is one of the topics that has attracted the attention of many researchers. Choosing the proper electrolyte for supercapacitors is one of the most significant factors affecting the performance of supercapacitors. In this paper, two classes of electrolytes, i.e. liquid electrolyte (ionic liquid electrolyte) and solid electrolyte (polymer electrolyte) are compared by molecular dynamics simulation. We consider the polymer electrolyte in linear and network configurations. The results show that although ionic liquid-based supercapacitors have a larger differential capacitance, since they have a smaller operation voltage, the amount of energy stored is less than polymer electrolyte-based supercapacitors. Also, our investigations indicate that polymer electrolyte-based supercapacitors have more mechanical stability. Therefore, they can be considered a very suitable alternative to liquid electrolyte-based supercapacitors that do not have known liquid electrolyte problems and display better performance.
\end{abstract}

%%%%%%%%%%%%%%%%%%%%%%%%%%%%%%%%%%%%%%%%%%%%%%%%%%%%%%%%%%%%%%%%%%%%%
%% Start the main part of the manuscript here.
%%%%%%%%%%%%%%%%%%%%%%%%%%%%%%%%%%%%%%%%%%%%%%%%%%%%%%%%%%%%%%%%%%%%%
\section{Introduction}
Supercapacitors (SCs), also known as Electric Double Layer Capacitors (EDLCs), have recently attracted much attention in the field of electrical energy storage. The SCs fill the gap between batteries and conventional capacitors in terms of energy and power density. They consist of two porous electrodes immersed in an electrolyte. Due to the potential difference between the electrodes, the charged electrodes repel the co-ions in the electrolyte while attracting their counter-ions, resulting in charge separation and charge storage. 

The SCs have higher energy density in comparison with conventional capacitors due to their porous electrodes with large surface areas and small charge separation distances \cite{sharma2020current}.
Also, compared to batteries, SCs have the advantages of higher power density induced by a fast charging/discharging rate (in seconds), a long cycle life (4,100,000 cycles), and high power density \cite{sharma2020current}. 
Despite their higher power density, they cannot store the same amount of energy as batteries \cite{sharma2020current}. Extensive efforts and research have been devoted to increasing the energy density of SCs to 20-30 Wh/L to solve the problems and satisfy the performance demands \cite{zhong2015review,naoi2012second,lu2013supercapacitors}. According to relation $E=\frac{1}{2} C V^{2}$, the energy density $(E)$ of SCs is proportional to the capacitance $(C)$ and the square of the voltage $(V)$. Therefore,  increasing either the capacitance or the voltage of a cell can be an effective way to achieve high energy density \cite{zhong2015review}.

The efficiency of SCs depends mainly on both electrolyte and electrode structure. The pore size and surface area of electrodes, ionic conductivity, and electrolyte operating voltage window play a significant role in developing high-performance and flexible SCs \cite{alipoori2020review}. Especially in the case of liquid electrolytes, they have some disadvantages for use in flexible SCs like being toxic and corrosive \cite{zhong2015review}, requiring high-cost packaging to fabricate, and are associated with leakage problems \cite{zhong2015review,zhang2009carbon}. In general, the critical features of an ideal electrolyte include: (1) a wide voltage and temperature window; (2) a high ionic conductivity; (3) a high chemical and mechanical stability; (4) well-matched with the electrolyte materials; (5) low volatility and flammability; (6) safety; and (7) simple processing with low cost \cite{du2020poly,zhang2009carbon,shaplov2015recent}. 

To overcome the limitations of liquid electrolytes, polymer electrolytes (PEs) were introduced.
In 1970, Armand first used PE in Lithium Ion Batteries (LIBs) and proposed LIBs with improved efficiency and energy density \cite{arya2017polymer}. PEs consist of a macromolecule matrix dissolved in a low viscosity and high dielectric constant organic solvent \cite{arya2017polymer}. PEs have many advantages such as avoiding liquid leakage and corrosion problems, good ionic conductivity, high chemical and mechanical stability, high energy density, safety, solvent-free condition, being light in weight, low cost, and simple manufacturing process \cite{du2020poly,arya2017polymer,zhang2009carbon,shaplov2015recent}. 

Due to the advantages of PEs, they are ideal candidates for use in SCs as electrolytes.
PEs for SCs can be classified into three categories: (1) solid polymer electrolytes (SPEs), (2) gel polymer electrolytes (GPEs), and (3) polyelectrolytes. The SPE is composed of a polymer (e.g., PEO) and a salt (e.g., LiCl), without any solvents. The ions in the SPE are transported through the polymer \cite{zhong2015review,alipoori2020review} and the polymer works as a host matrix for ion movement \cite{zhong2015review,aziz2018conceptual}. In contrast, the GPE consists of a polymer host (e.g., PVA) and a liquid electrolyte or a conducting salt dissolved in a solvent \cite{zhong2015review,lin2012cation}. The polymer in GPE is swollen by the solvent and acts as a dynamic moving matrix. The conductivity of ions occurs through the solvent instead of the polymer phase \cite{zhong2015review,lin2012cation}. In GPEs, the liquid electrolyte generally provides free ions that participate in conductivity enhancement and also acts as a conductive medium. In addition, the polymer provides perfect mechanical stability by increasing the viscosity of the electrolyte \cite{alipoori2020review}. Recently, researchers have shown that using Ionic Liquids (ILs) can improve ionic conductivity and cell voltage, resulting to improvement in the electrochemical performance of GPE \cite{alipoori2020review}. By its softening effect on the polymer chains, IL can increase the electrolyte's ionic conductivity and facilitate ion transfer \cite{alipoori2020review}. GPE based on ILs and linear polymers usually exhibit poor mechanical properties, including both strength and flexibility, because of their few polymer chain entanglements separated by small molecules \cite{taghavikish2018poly}. To generate polymer networks, cross-linking strategies have been proposed in recent years. The behaviors and performance of polymer gels are largely determined by the structure of the polymer network that makes up the gel. This is due to the interaction between the network and the solvent \cite{alipoori2020review}. Gels generally have high mobility because the polymer networks are dissolved by a large amount of entrapped solvent \cite{alipoori2020review}.

In the polyelectrolyte, ionic conductivity is created by charged polymer chains \cite{taghavikish2018poly}. As it turns out, each type of these solid-state electrolytes has its advantages and disadvantages. Typically, GPEs have the highest ionic conductivity among the three types of solid-state electrolytes \cite{alipoori2020review,taghavikish2018poly}. Due to the liquid phase in the GPE, its ionic conductivity is significantly higher than dry SPE. Therefore, GPE-based SCs currently dominate the products of solid electrolyte-based SCs. Several polymer matrices have been explored for preparing GPEs in the role of host polymer including: poly (vinyl alcohol) (PVA), poly (acrylic acid) (PAA), potassium polyacrylate (PAAK), poly (ethyl oxide) (PEO), poly (methylmethacrylate) (PMMA), poly (ether ether ketone) (PEEK), and poly (vinylidene fluoride-co-hexafluoro-propylene) (PVDF-HFP) \cite{zhong2015review}. 

Beyond all of the experimental achievements \cite{liew2014good,zhang2009carbon,karaman2018enhanced,kumar2012gel,pandey2010performance}, modeling the EDLCs under different physical conditions would provide a lot of insight into the system's physics. Modeling the microstructure will reveal how the related dynamics for charge carriers happen. So far, numerous types of research have been done on liquid electrolyte-based SCs to explore the performance of different kinds of liquid electrolyte-based SCs and the effect of electrode structure and its pore size \cite{breitsprecher2014coarse,breitsprecher2017effect,paek2015influence,yang2017molecular}.    
Our previous work \cite{eyvazi2022molecular}, models the third classification of polymer electrolytes (polyelectrolytes). In this work, considering the advantages of GPEs compared to other polymer electrolytes and their useful applications, we investigated the behavior of GPE-based SCs using the molecular dynamics simulation method. Here, we compared IL-based and GPE-based SCs in order to find out the effects of adding host polymers to the IL electrolyte. In the first section, we describe the method of simulation under different conditions. Based on the results we get from our model, we discuss the pros and cons of using GPEs and IL electrolytes in SCs.

\section{Method}
\subsection{Systems' model and force field parameters}
Molecular Dynamics (MD) simulations have been widely used to describe the behavior of SCs. Using MD simulations, we compared liquid electrolyte-based SCs with PE-based SCs under various conditions. The simulations were carried out by the CAVIAR software package \cite{biagooi2020caviar}. We have investigated the performance of three different systems with similar electrodes but different electrolytes. The electrolytes were confined between two electrodes with single slit-pore geometry, placed at a distance of 15 nm from each other. This model called a slit-pore model was introduced by Breitsprecher et al \cite{breitsprecher2017effect,breitsprecher2018charge} for simulating porous media.

The slit-pore length was set to 15 nm, in order to be compared with the bulk region and the width of the pore is 2.5 nm (larger than the size of two ionic particles) so that at least two ionic particles can pass through the pore \cite{kondrat2011superionic}.

To compare the electrolyte role we have simulated the above mentioned pore structure with different electrolyte systems; liquid electrolyte (system A), linear polymer electrolyte (system B), and polymer network electrolyte (system C). By comparing the results, we can discuss the effects and the function of different electrolytes.
\subsubsection{System A: Liquid electrolyte-based supercapacitors}
Our first model contains IL electrolytes confined between those described electrodes. We used Coarse-Grained (CG) models of ILs, where ions are soft spheres with diameters d\textsubscript{cation} = d\textsubscript{anion} = 1 nm and valency q = $\pm$ 1, in units of the elementary charge. The mass of cations and anions are m\textsubscript{cation} = $117.17 \frac{g}{mol}$ and m\textsubscript{anion} = $86.81 \frac{g}{mol}$  corresponding to EMIM$^\textmd{+}$ and BF4$^\textmd{-}$.

All the particles interact with the non-bonding Lennard-Jones (LJ) potential:
\begin{equation}
	\centering
	V_{LJ}(r_{ij}) = \left\{
	\begin{array}{lr}
		4 \varepsilon_{ij} [(\frac{\sigma_{ij}}{r_{ij}})^{12} - (\frac{\sigma_{ij}}{r_{ij}})^6]& ; \quad r \leq r_{cut} \\
		0 & ; \quad r \geq r_{cut}
	\end{array} \right.
	\label{eq_lj}
\end{equation}
Where $r_{ij}$ is the relative distance between each pair of particles, $\varepsilon$ and $\sigma$ are the length and energy parameters. In calculating the LJ potential, $r_{cut} = \sqrt[6]{2}~\sigma$ is the cutoff length to ensure short-range repulsive force at any distance. 
In addition, charged particles also interact through the electrostatic non-bonding potential:
\begin{equation}
	\centering
	V_C(r_{ij}) = \sum_{i}\sum_{i<j}\frac{z_{i}z_{j}e^{2}}{4\pi \epsilon |\textbf{r}_{i} - \textbf{r}_{j}|}
	\label{eq_coul}
\end{equation}
The $\epsilon$ and $e$ are the electric permittivity and the electrical charge. The $z = \pm 1$ is the valency of ions.  For the electrostatic interactions we used the electric permittivity of $\epsilon = 10$, a typical value for ILs at $300$ $^{\circ}K$ \cite{breitsprecher2014coarse}.

\subsubsection{System B: Linear polymer electrolyte-based supercapacitors}
\begin{figure}[h!]
	\centering
	\includegraphics[width=0.85\linewidth]{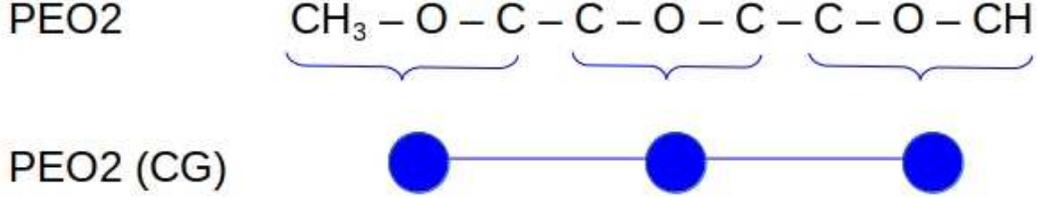}
	\caption{PEO2 and its CG model.}
	\label{fig_PEO}
\end{figure}

In system B, we added polymer hosts to system A to simulate GPE-based SCs. By comparing these two systems' behavior, we can find out the advantages of one over another. 
Poly(ethylene oxide) (PEO) with the formula H$_3$C-O-(CH$_2$-CH$_2$-O)$_n$-CH$_3$ \cite{lee2009coarse}, the most common polymer with a broad range of applications in polymer chemistry \cite{croce1998nanocomposite}, biotechnology \cite{allen1991liposomes}, and medical science \cite{arya2017polymer,ma2020coarse}, is used as a host polymer in our simulations. 
We used MARTINI-like CG models for PEO simulation \cite{ma2020coarse,lee2009coarse}. Fig.\ref{fig_PEO} represents the CG model for PEO2. 

Generally, the MARTINI potential consists of bond, angle, LJ, electrostatic, and torsional terms. In our system, two consecutive beads in each chain are connected by a harmonic stretching spring whose potential is taken to be:
\begin{equation}
	V_{bond}(r) = \frac{1}{2}K_b (r - r_0)^2
	\label{eq_bond}
\end{equation}
Where $K_b$ is the bond force constant, $r$ is the instantaneous bond length, and $r_0$ is the equilibrium length of the bond. Bond angle is the angle formed between three atoms across at least two bonds which can be described as:
\begin{equation}
	V_{angle} (\theta) = \frac{1}{2} K_{\theta} (cos \theta - cos \theta_0)
	\label{eq_angle}
\end{equation}
and the torsion angle (also called dihedral angle) is defined by 3 consecutive bonds involving 4 atoms: 
\begin{equation}
	V_{dihedral} (\phi) = \sum_{n=1}^{4} K_{\phi , n} (1+cos(n \phi - \phi_n))
	\label{eq_dihedral}
\end{equation}
Where $n$ and $\phi$ are the multiplicities and offsets of the n individual dihedral terms. The force constants for these bonded interactions are listed in Table.\ref{table_1} \cite{lee2009coarse}.
{\renewcommand{\arraystretch}{1.5}
	\begin{table*}[!ht]
		\centering
		\begin{tabular}{|cc|cc|ccc|}
			\hline
			\hspace{5pt} Bond & &\hspace{5pt} Angle & & \hspace{5pt} Dihedral & &\hspace{5pt}\\
			\hline
			$r_0$({\AA}) & $K_b$($\frac{kJ}{mol~nm^2}$) & $\theta_0$ ($deg$)& $K_{\theta}$ ($\frac{kJ}{mol}$)& $\phi_n$ ($deg$) & $K_{\phi}$ ($\frac{kJ}{mol}$) & n \\
			\hline
			3.30 & 17000 & 130 & 85 & 180 & 1.96 & 1 \\
			&       &     &    & 0   & 0.18 & 2 \\
			&       &     &    & 0   & 0.33 & 3 \\
			&       &     &    & 0   & 0.12 & 4 \\
			\hline		
		\end{tabular}
		\caption{Parameters for bonded interactions for modeling CG PEO.}
		\label{table_1}
	\end{table*}
	
	The LJ parameters in Eq.\ref{eq_lj} for CG PEO beads were set to $\sigma_{ij}=5$ {\AA} and $\varepsilon_{ij}=3.375 ~\frac{kJ}{mol}$. 
	Each chain in CG PEO modeling has 25 spherical beads connected with the mass $\textmd{m}_{\textmd{PEO}} = 60.05376~\frac{g}{mol}$. 
	
	\subsubsection{System C: Network polymer electrolyte-based supercapacitors}
	Nowadays,  polymer networks have received much attention due to their characteristics and have been used in various applications \cite{wu2012langevin}. In network polymer structures,  all polymer chains are directly or indirectly linked to each other.
	The cross-linking of polymer chains into complex networks is a promising strategy to improve the mechanical strength of a GPE and provide dimensional stability at high temperatures \cite{lehmann2020well} which has been widely investigated \cite{wu2012langevin}.  As a model for the polymer network, we consider a cross-link of linear chains as follows. First, we highlighted some monomers as cross-links. As the cross-links are closer than the distance $r_0 = 3.30 $ {\AA}, a new permanent bond is formed between the cross-linkers, acting like cross-linked monomers \cite{wu2012langevin}. Once equilibrium is reached, the polymer network can be used instead of linear polymers in our simulations. Fig. \ref{fig_lpoly} demonstrates the linear polymers we consider in our simulation as system B and Fig. \ref{fig_npoly} displays cross-linked polymers in system C.
	\begin{figure}[h]
		\centering
		\begin{subfigure}{0.4\linewidth}
			\centering
			\includegraphics[width=\linewidth]{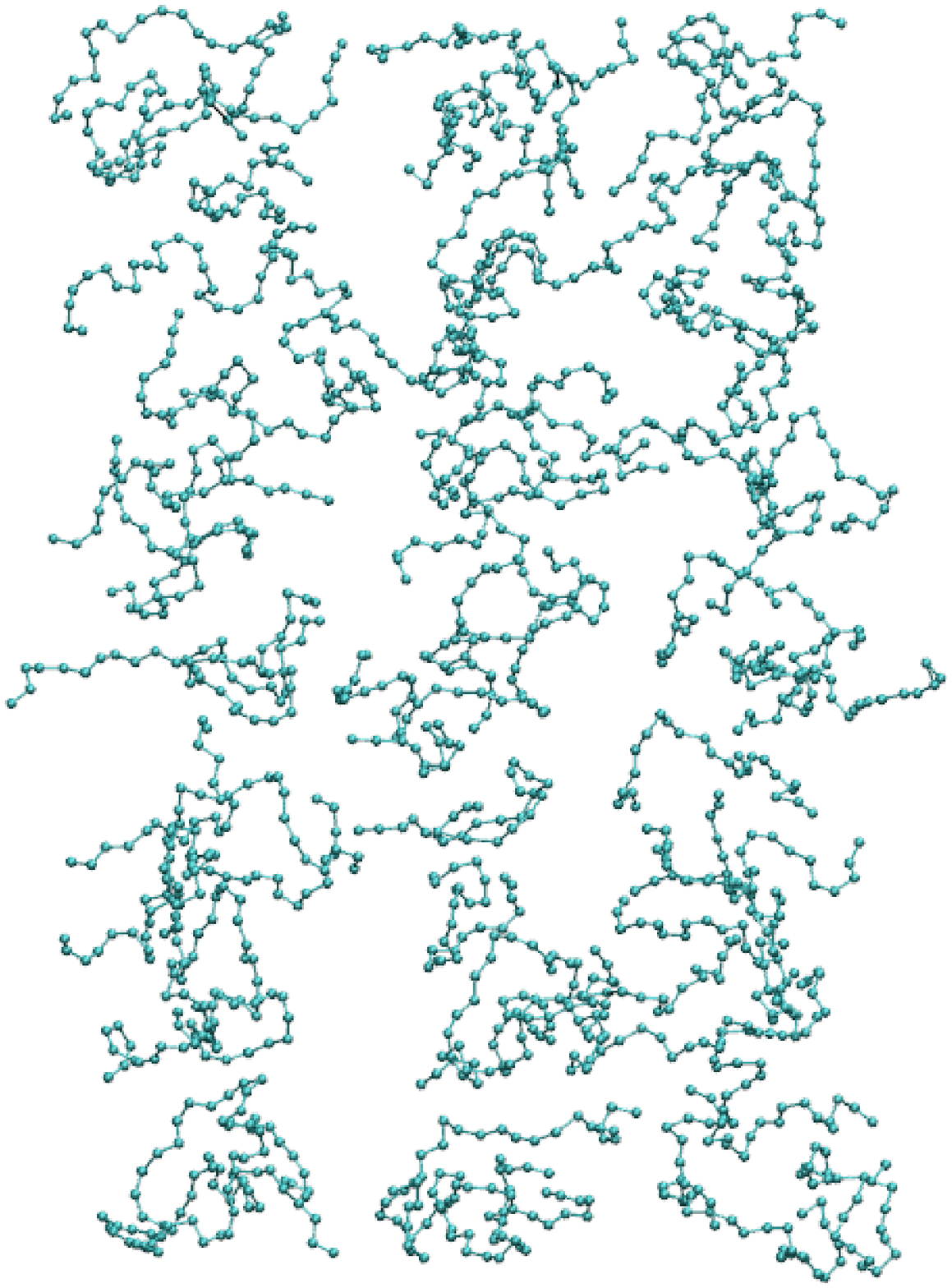}
			\caption{}
			\label{fig_lpoly}
		\end{subfigure}
		\hfill
		\begin{subfigure}{0.4\linewidth}
			\centering
			\includegraphics[width=\linewidth]{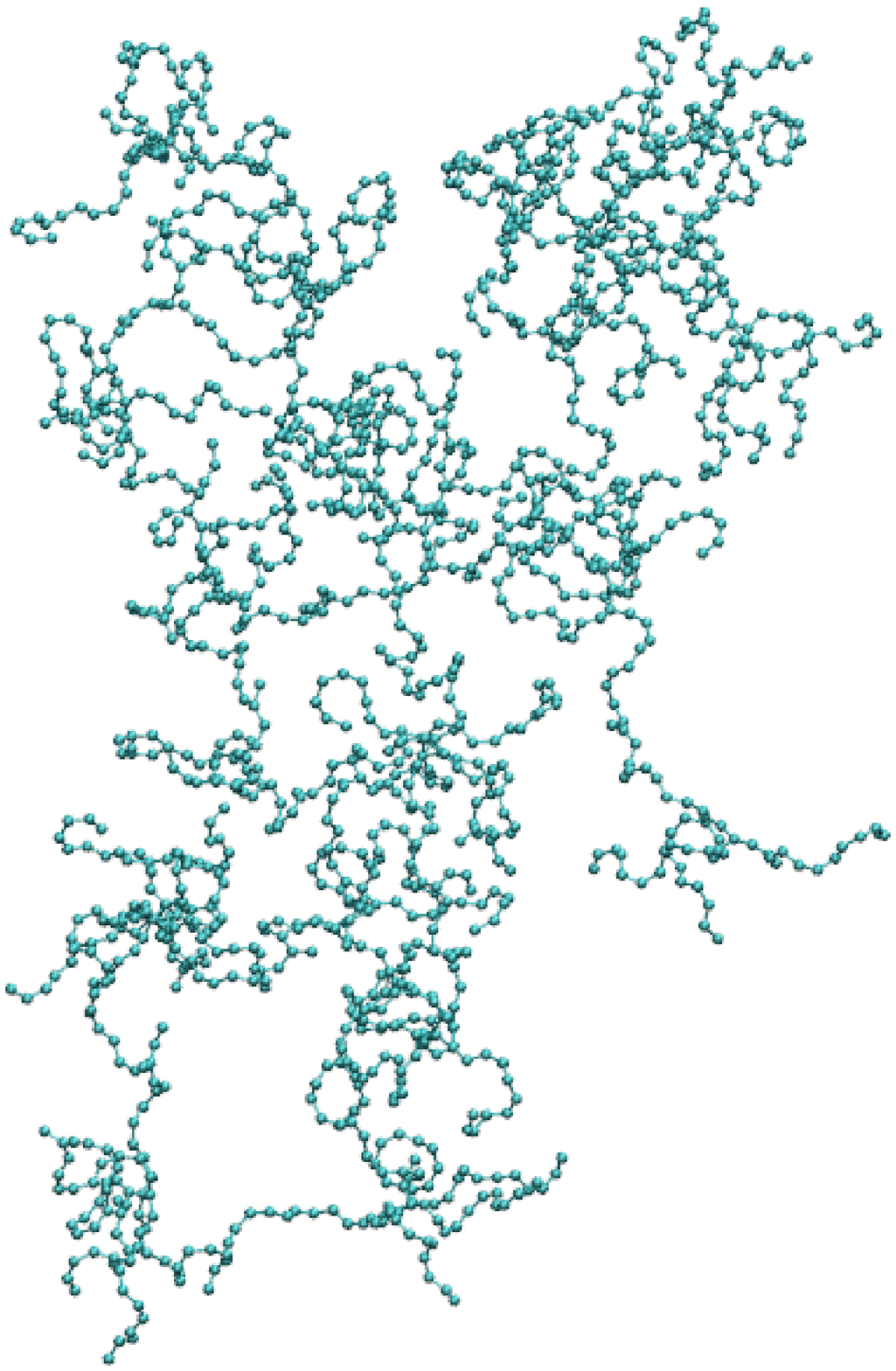}
			\caption{}
			\label{fig_npoly}
		\end{subfigure}
		\caption{(a) Linear polymers distribution in system B and (b) Cross-linked polymers in system C.}
		\label{fig_poly}
	\end{figure} 
	
	\subsection{Simulation details and electrodes model:}
	The simulations were performed using the Langevin thermostat to keep the temperature constant. Thus, the equation of motion of the system was calculated by the Langevin relation:
	\begin{equation}
		\centering
		m \ddot{\textbf{r}}_{i} = -\nabla_{i} U(\{\ \textbf{r}_{j}(t)\})  - \gamma m \dot{\textbf{r}}_{i} + \textbf{F}_{i}.
		\label{eq_langevin}
	\end{equation}
	The first term describes the deterministic forces between particles (the force acting on atom $i$ due to the interaction potentials), and the last two terms implicitly consider the effect of the solvent by coupling the system to a Langevin thermostat which maintains a constant average temperature of the system. The parameter $\gamma$ is the friction coefficient and $\textbf{F}_{i}$ is a Gaussian distributed random force with \cite{kotelyanskii2004simulation}
	\begin{equation}
		\centering
		\begin{array}{lr}
			\langle \textbf{F}_{i}(t) \rangle = 0 \\ 
			\langle \textbf{F}_{i}(t) \textbf{F}_{j}(t^{\prime}) \rangle = 6k_{B}Tm \gamma \delta_{ij} \delta(t-t^{\prime}).
		\end{array}
		\label{eq_langevinDist}
	\end{equation}
	
Our simulation box included 630 charged particles, where half of which are cations and the remains are anions, and 1575 monomers. The density of the bulk region was set to $1.07 \frac{1}{nm^3}$ and the pore size in three defined systems was equal.
To simplify the units, the reduced LJ unit was used.
The length was scaled with ion size $\tilde{l}=\sigma=5$ {\AA}, and the mass unit was $\tilde{m}=144~\frac{g}{mol}$. The energy unit and the charge unit were $\tilde{\varepsilon}=1~\frac{kJ}{mol}$ and $\tilde{q}=e$. By applying $\tilde{t}=\sqrt{\dfrac{\tilde{m} \sigma^{2}}{\tilde{\varepsilon}}}$, the time unit was obtained $\tilde{t}=5~ps$. In addition, temperature and voltage were scaled as $\tilde{T}=\dfrac{\tilde{\varepsilon}}{k_B}=120.267$ $^{\circ}K$ and $\tilde{V}=\dfrac{\tilde{q}}{\tilde{\varepsilon}}=0.01036~V$.
Temperature was $k_{B}T = 2.5~ \varepsilon$ which corresponded to $300$ $^{\circ}K$ in real unit. The friction coefficient in the Langevin equation was set to $\frac{1}{\gamma} = \frac{1}{\tilde{t}}$ and the time step was $\Delta t = 0.0005~\tilde{t}$ equal to $6~fs$.
	
	The electrodes were built from carbon atoms with the following parameters: $\sigma_{C} = 3.37$ {\AA} and $\varepsilon_{C} = 1~\frac{kJ}{mol}$. Here, using the Poisson to Laplace Transformation (PLT) method, which is recently developed in the CAVIAR package \cite{biagooi2020caviar}, the electrodes are surfaces with a constant potential. Periodic boundary condition applied in the XY plane and the long-range coulombic interaction performed using a 1D Ewald algorithm \cite{lee2009ewald} with $R_C=15~\sigma$ as the cutoff distance for electrostatic interaction. The systems were simulated at 17 various potential differences between electrodes: 0, 0.25, 0.5, 0.75, 1, 1.25, 1.5, 1.75, 2, 2.25, 2.5, 2.75, 3, 3.25, 3.5, 3.75, 4 V. Calculations were done after the system reached equilibrium.
	
	\section{Results and Discussion}
	We begin our discussion with a snapshot of three defined systems. Fig.\ref{fig_structure} illustrates a snapshot from cross-section of the systems configuration and ions separation in the equilibrium condition for $\Delta V=2~V$. In order to find out the charge separation process and the behavior of these three systems, it is necessary to conduct numerous calculations and measurements. We will discuss this below. 
	  
	\begin{figure*}[ht]
		\centering
		\begin{subfigure}{0.45\linewidth}
			\centering

			\includegraphics[width=0.85\textwidth]{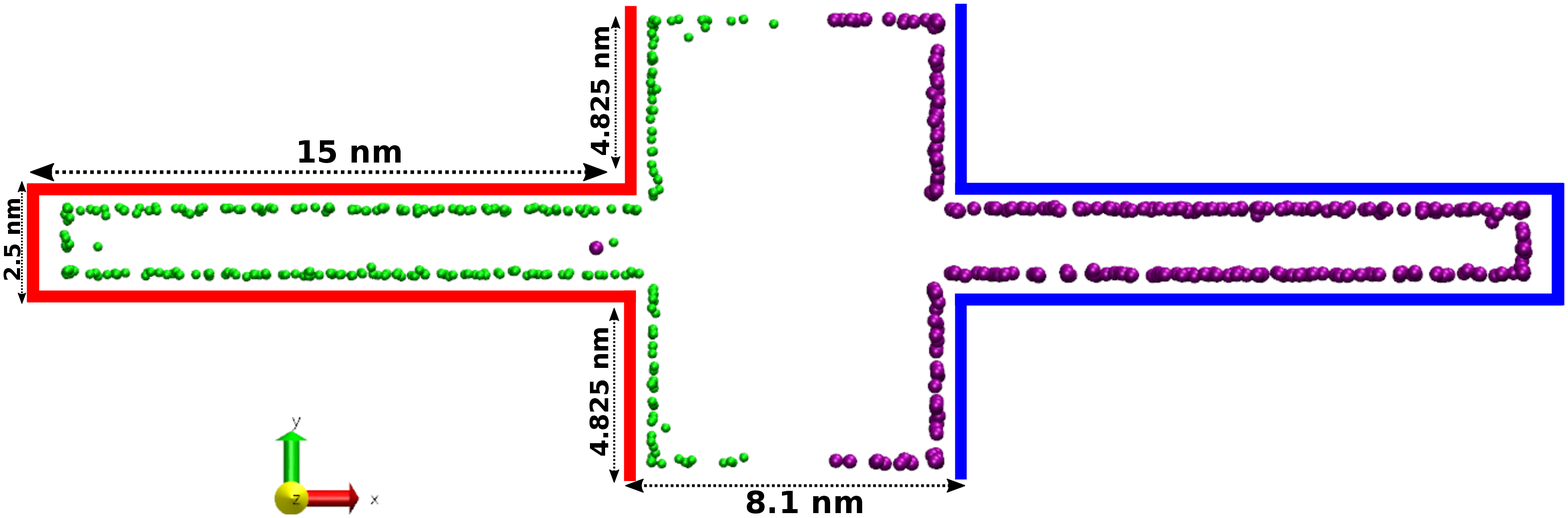}
			\caption{}
			\label{fig_ilstructure}
		\end{subfigure}
		\vfill
		\begin{subfigure}{0.45\linewidth}
			\centering
			\includegraphics[width=\textwidth]{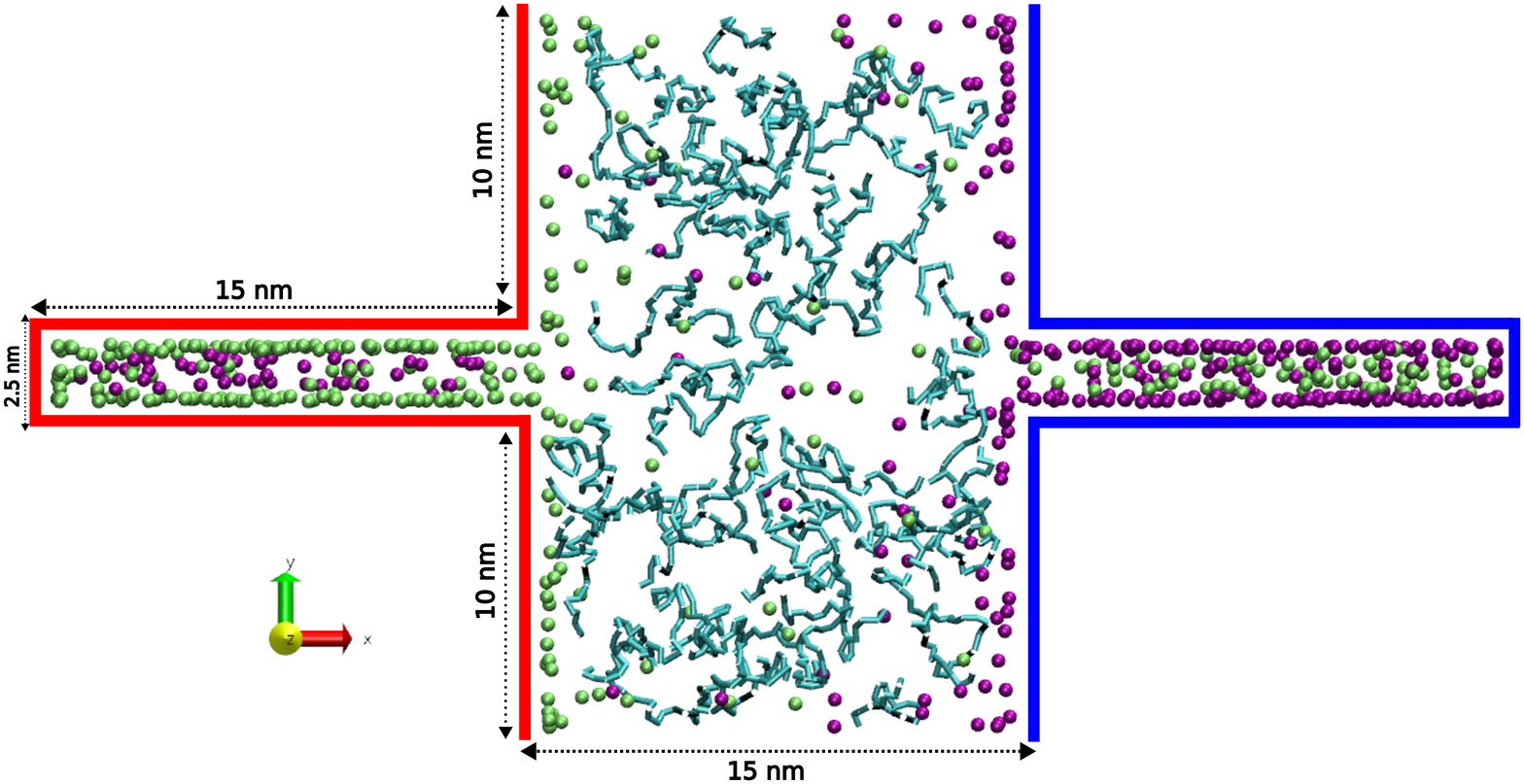}
			\caption{}
			\label{fig_lstructure}
		\end{subfigure}
		\hfill
		\begin{subfigure}{0.45\linewidth}
			\centering
			\includegraphics[width=\textwidth]{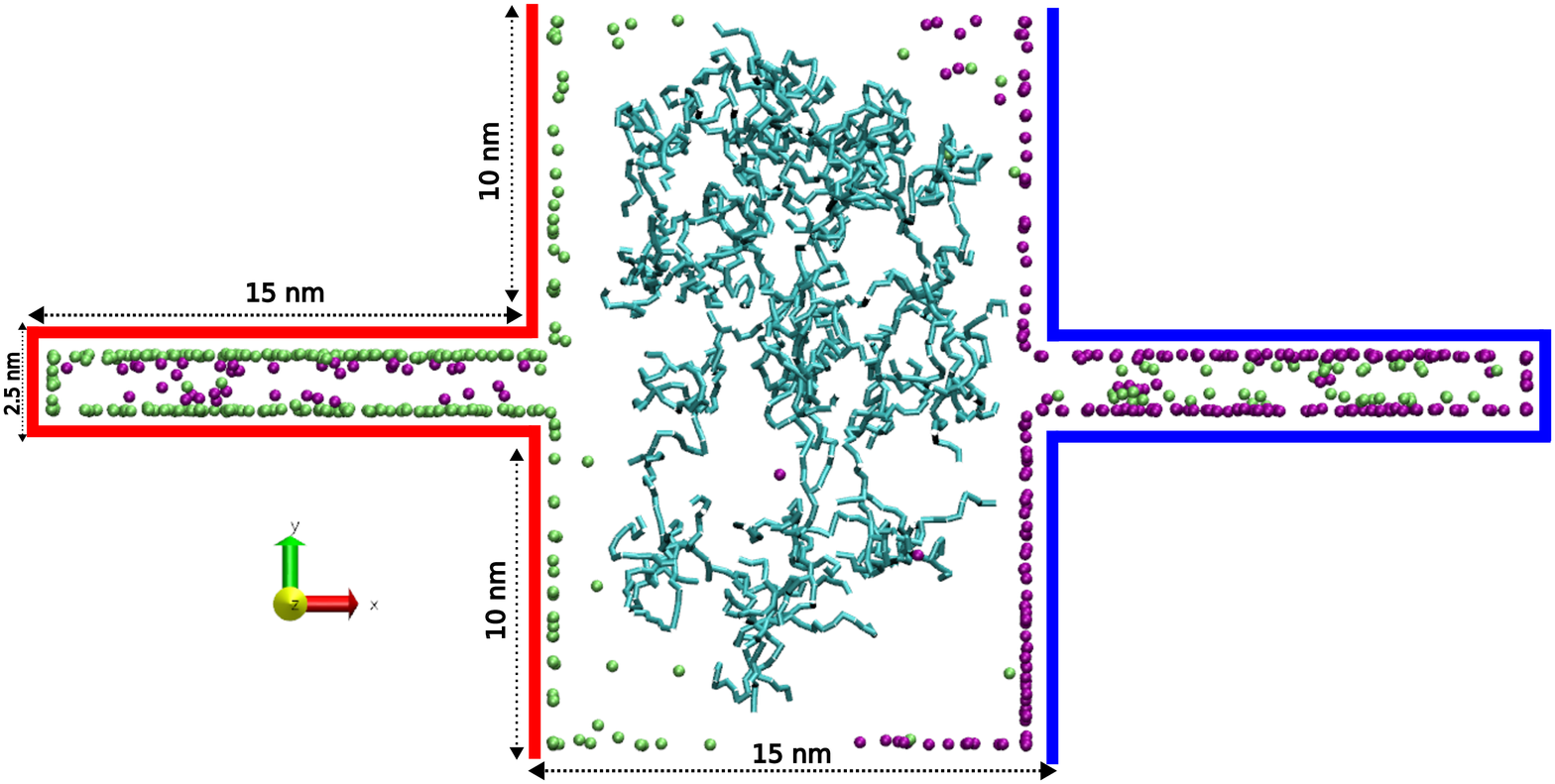}
			\caption{}
			\label{fig_nstructure}
		\end{subfigure}
		\caption{The systems configuration at $\Delta V=2V$ after equilibration for (a) System A (b) System B and (c) System C.}
		\label{fig_structure}
	\end{figure*}
	
	\subsection{Charging and Differential Capacitance}
	Initially, we discuss the results of the charging process. Fig.\ref{fig_MeanCharge} demonstrates average charge density accumulation on the surface of the electrodes for the three systems as a function of the applied potential difference between the electrodes. According to the figure, increasing the potential difference results in a high concentration of counterions on the electrodes' surface. Therefore average induced charge grows by increasing the applied potential until reaching saturation. For system A, the induced charge is greater than in the two other systems. Based on the figure, the saturation voltages for system A is $\Delta V=2~V$, and for systems B and C is $\Delta V=3.5~V$. Since IL-based SCs are saturated at a lower voltage than polymer-based SCs, their operating voltage window is also smaller.  
	
	For systems B and C, the induced charges on the electrodes for voltages $\Delta V < 1.25~V$, are almost the same. In voltages, $\Delta V > 1.25~V$, induced charges in system C, are greater than system B until both systems reach saturation in $\Delta V=3.5~V$.  
	
	\begin{figure}[ht]
		\centering
		\includegraphics[width=0.9\linewidth]{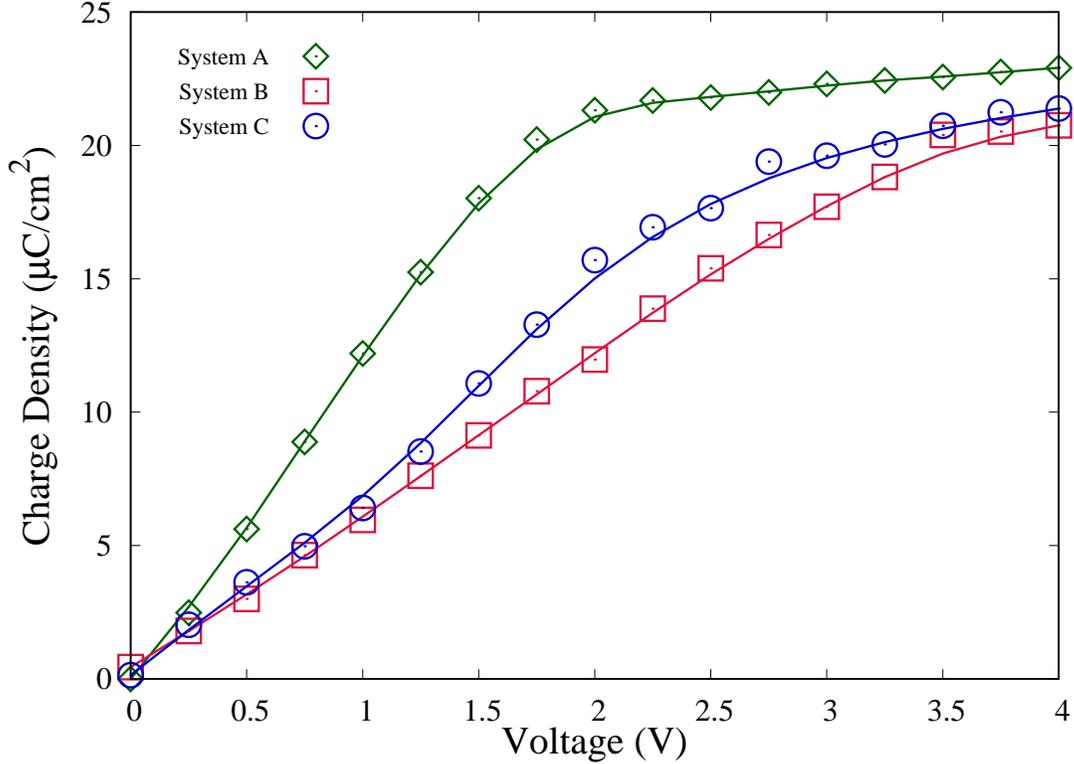}
		\caption{Mean induced charge density on the electrodes surface. The points in the figure display the simulation data and the solid lines in the plots represent the smoothed data.}
		\label{fig_MeanCharge}	
	\end{figure}

	On the other hand, charge storage in SCs is described by the critical quantities of the Differential Capacitance, $C_d(V)$, and the energy density. The $C_d(V)$ is the derivative of the electrodes surface charge with respect to the applied potential difference between the electrodes \cite{bossa2018modeling}:
	\begin{equation}
		C_d(V)=\frac{dq}{dV}.
		\label{eq_dc}
	\end{equation}
	According to the Eq.\ref{eq_dc}, the $C_d(V)$ is the derivative of the curves in Fig.\ref{fig_MeanCharge}. Fig.\ref{fig_dc}, which is obtained by the numeric derivation of smooth curves of $q(V)$, shows the $C_d(V)$ plots for three systems.
	
	\begin{figure}[ht]
		\centering
		\includegraphics[width=0.9\linewidth]{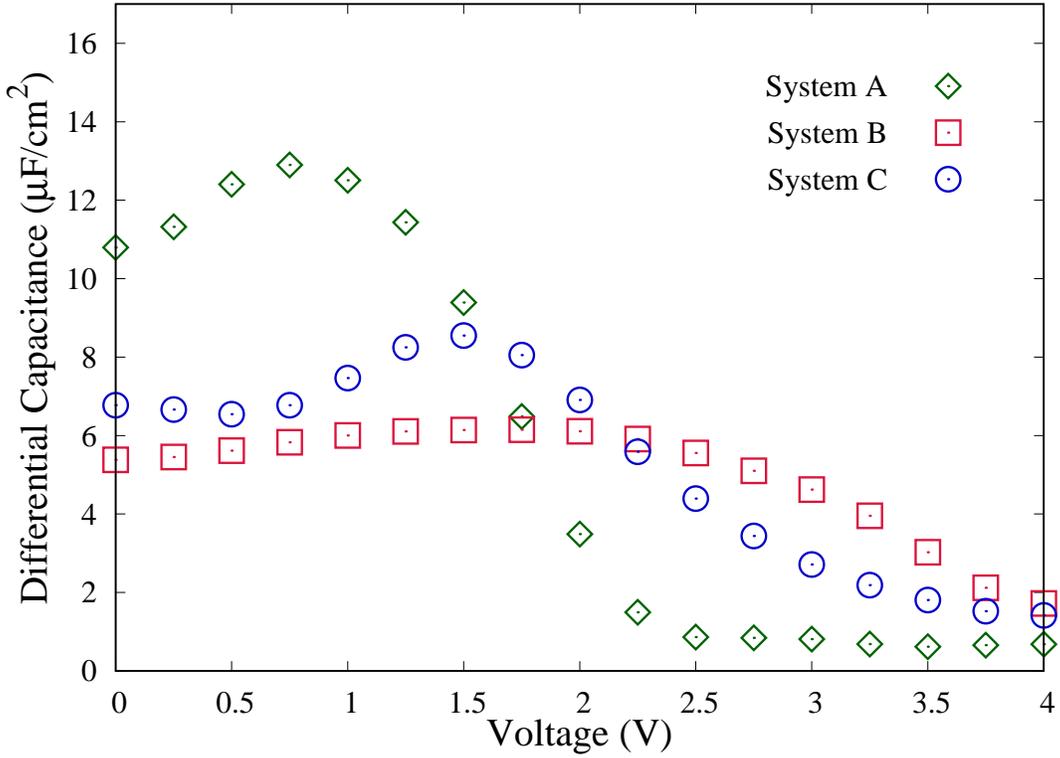}
		\caption{Differential Capacitance for three systems.}
		\label{fig_dc}
	\end{figure}
	
	Based on Kornyshev theory for most ILs, $C_d(V)$ displays a bell-shape, with a maximum at the Potential of Zero Charge (PZC) or a camel-shape with two peaks.
	In both cases, $C_d(V)$ decreases at large potential difference. The reason is, the ions are only allowed to pack up to a given maximum density in the double layer. Therefore, by increasing the potential difference, the ions concentration near the electrodes' surface reaches their maximum value and the effective diffuse layer thickness actually grows larger, leading to a decrease in $C_d(V)$ \cite{frischknecht2014electrical,frischknecht2014electrical}. Fig.\ref{fig_dc} is a plot of the $C_d(V)$ for three systems, which display camel-shaped figures. 
	There is limited change in the $C_d(V)$ curve for system B, indicating the surface charge density changes less rapidly in system B with increasing potential. In contrast, there is a peak at $\Delta V=1.5~V$ in the $C_d(V)$ plot of system C which points out that more counter-ions can be condensed on the electrodes' surface of system C in comparison with system B.
	The $C_d(V)$ peak for system A is the highest among systems B and C. 
	The result of Korneyshev theory is observed in Fig.\ref{fig_MeanCharge} and Fig.\ref{fig_dc}. 
	In systems A, B and C, the saturation voltage is 2, 3 and 3.5 $V$. Due to this, in the $C_d(V)$ plot, the DC drops at voltages above these values.
	
	As mentioned before, energy density is another useful parameter that shows the efficiency of SCs. SC with a higher energy density is more efficient in electrical devices.
	The stored energy density in SCs is obtained \cite{bi2020molecular}:
	\begin{equation}
		E(V)=\int_{0}^{V} V~C_d(V)~dV
		\label{eq_energy}
	\end{equation}
	Based on the above equation, the plot of energy density for the three systems is calculated and plotted in Fig.\ref{fig_energy}. 
	According to this plot, for $\Delta V < 2~V$ the energy density of system A is greater than systems B and C. For $\Delta V > 2~V$ systems B and C have higher energy density which indicates that the GPEs-based SC has more efficiency than IL-based SCs in higher potential difference.  
	\begin{figure}[ht]
		\centering
		\includegraphics[width=0.9\linewidth]{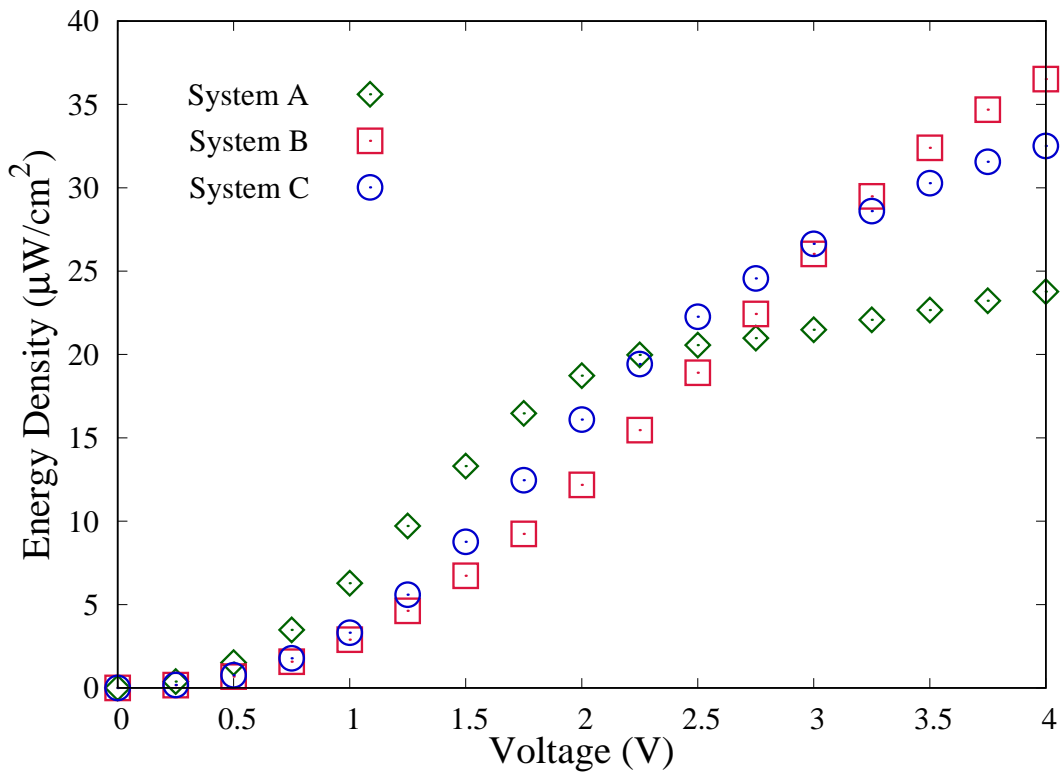}
		\caption{The stored energy density in three systems.}
		\label{fig_energy}
	\end{figure}
	
	\section{Structural investigation: Ion density profile }
	Further insight into the ion structure near the electrodes' surface is gained in this section.
	\begin{figure*}[ht]
		\centering
		\begin{subfigure}{0.3\linewidth}
			\centering
			\includegraphics[width=0.9\textwidth]{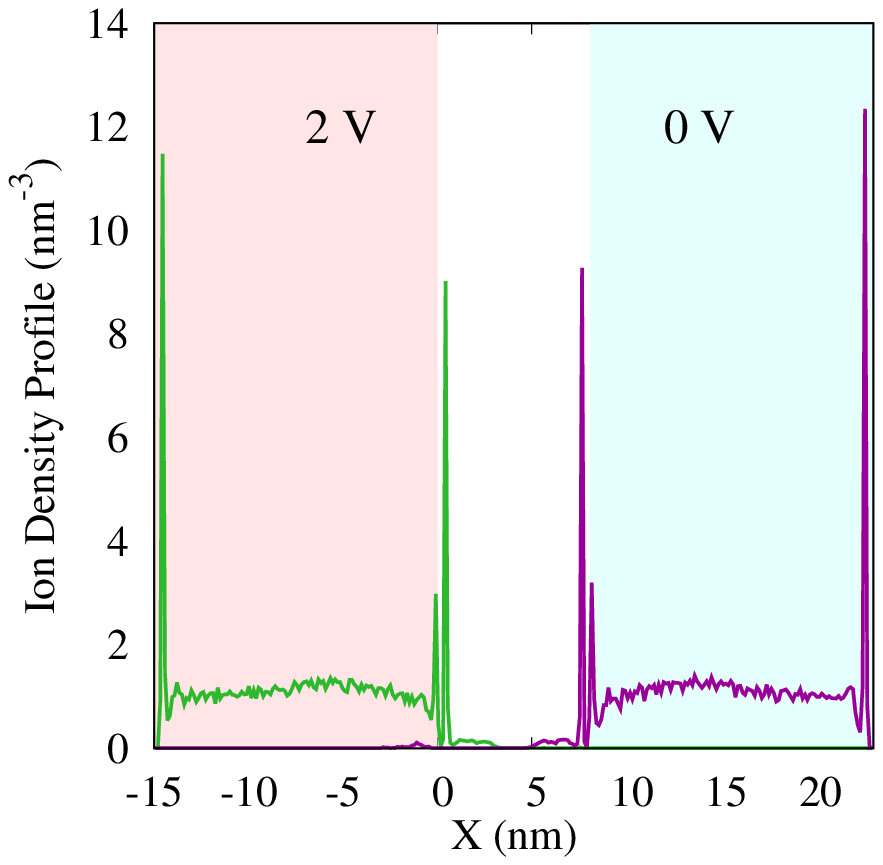}
			\caption{}
			\label{fig_profileA}
		\end{subfigure}
		\hfill
		\begin{subfigure}{0.3\linewidth}
			\centering
			\includegraphics[width=\textwidth]{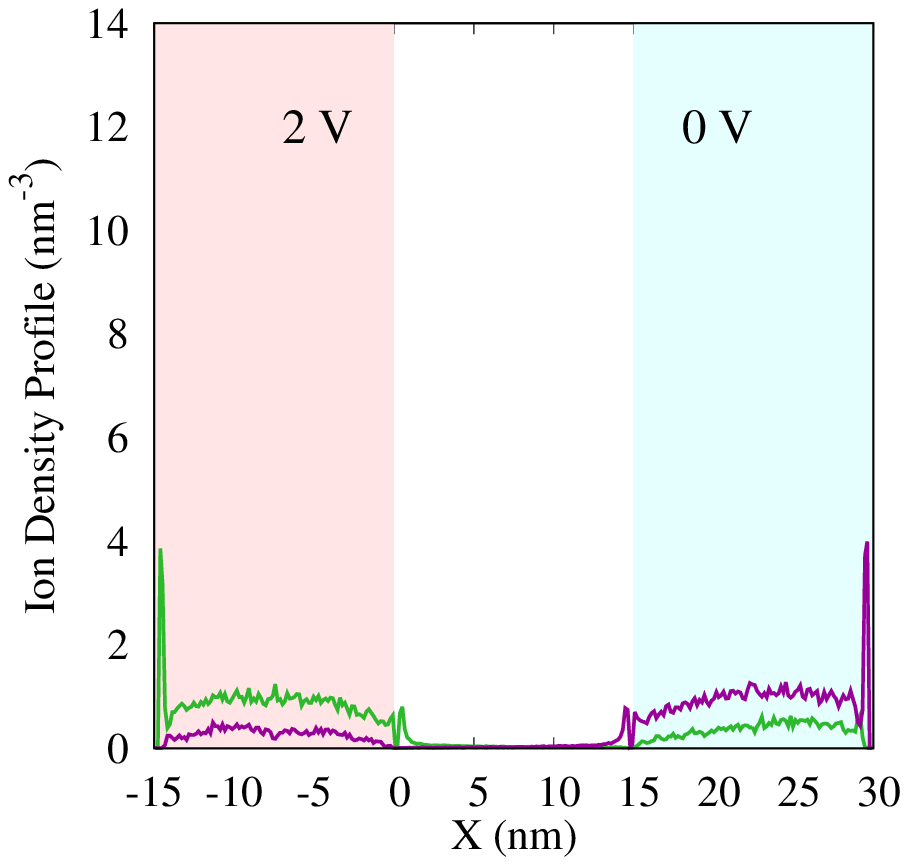}
			\caption{}
			\label{fig_profileB}
		\end{subfigure}
		\hfill
		\begin{subfigure}{0.3\linewidth}
			\centering
			\includegraphics[width=\textwidth]{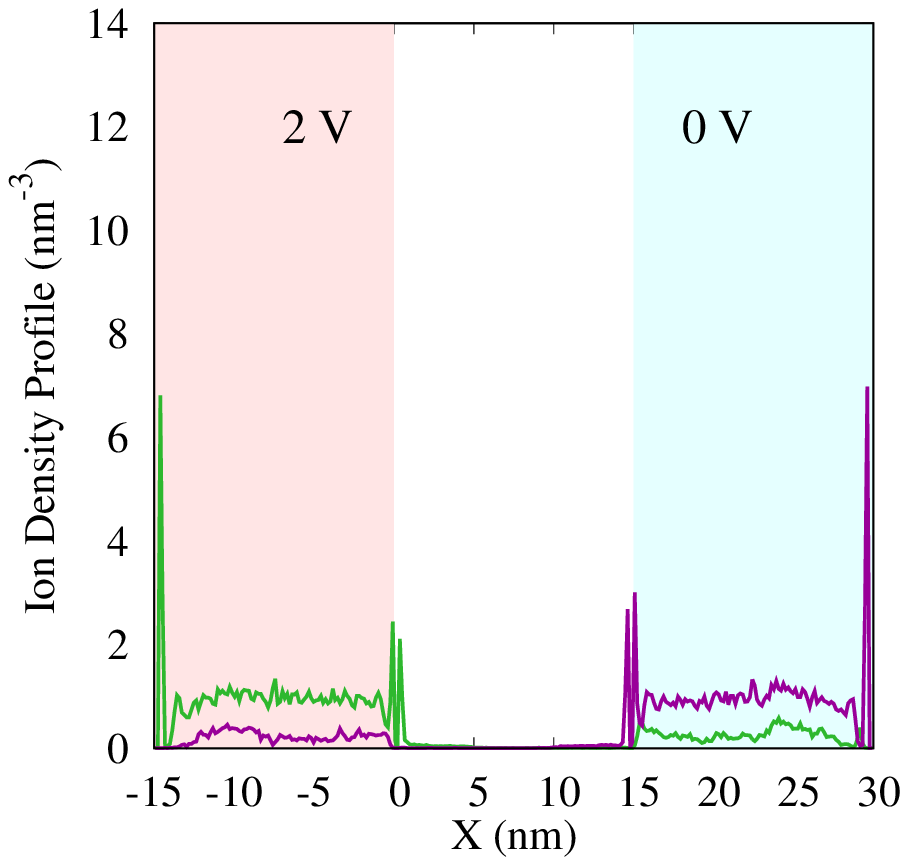}
			\caption{}
			\label{fig_profileC}
		\end{subfigure}
		\caption{Ion density profile of three systems along x axis for (a) System A (b) System B and (c) System C. Green and red graphs display positive and negative ions.}
		\label{fig_profile}
	\end{figure*}
	Due to the constant density in all three systems, system A has a smaller bulk region between its electrodes than systems B and C. 
	It is expected that the ion density profiles will oscillate near the charged surface in IL-based SCs \cite{frischknecht2014electrical}, as shown in Fig.\ref{fig_profileA} for $\Delta V=2~V$.
	Fig.\ref{fig_profileB} and Fig.\ref{fig_profileC} depict ion density profiles for systems B and C at $\Delta V=2~V$ after reaching equilibrium. Similar to ILs, the ion density profile of these systems exhibits layers and oscillations near electrode surfaces. While the conductive electrodes repel co-ions and attract counter-ions, co-ions can still be observed inside the pores. 
	The highest peak in Fig.\ref{fig_profileA} can be seen at the entrance and end of the pores, which contributes to the high induced charge on the surface of the electrodes.
	Fig.\ref{fig_profileC}, the ion density profile of system C, displays a higher peak at the entrance of pores in analogy with the profile of system B. This leads to more induced charge on the electrodes' surface and eventually higher energy density.
	
	The linear polymers can be placed near the electrode surfaces as well as in bulk space in system B, in contrast to system C, which places the polymer network only in bulk space. As a result, the accessible space for ions near electrode surfaces is reduced. Accordingly, system B exhibits a lower ion density at the pore entrance than systems A and C.
	
	\section{Dynamical investigation: Mean Squared Displacement and Diffusion coefficient}
	
	Gels with cross-linked polymers demonstrated higher mobility \cite{alipoori2020review}. Since the mobility of ions is proportional to the diffusion coefficient $(D)$ we need to calculate the diffusion coefficient.  
	The ionic diffusion coefficient is derived from the Mean-Squared
	Displacement (MSD) curve using the 3D diffusion relation: 
	\begin{equation}
		MSD \equiv <\Delta \textbf{r}(t)^2 > =\frac{1}{N}\sum_{i=1}^{N} <\textbf{r}_i(t)^2 - \textbf{r}_i(0)^2>,  
		\label{eq_MSD}
	\end{equation}
	where $N$ is the number of particles. Diffusion coefficient can be calculate using the following equation;
	\begin{equation}
		D = \lim_{t\rightarrow \infty} \dfrac{<\Delta \textbf{r}(t)^2 >}{6t}.
		\label{eq_diffusion}
	\end{equation}
	According to the above equation, also known as Einstein's relation \cite{son2020ion}, the self-diffusion of particles is calculated from the slope of the MSD curve over time.
	
	We calculated MSD for clusters of linear polymers in system B and cross-linked polymers in system C.
	According to Fig.\ref{fig_msdPEO}, MSD for linear polymers and cross-linked polymers display superdiffusive motion. On the other hand, by taking the slope of the MSD into account, the diffusion coefficient for cross-linked polymers is greater than linear polymers and therefore the mobility of cross-linked polymers is higher. This result can also be seen in Fig.\ref{fig_vmd}. In this figure, two snapshots of both polymers are shown for equal time intervals $\Delta t = 10$ timestep. Fig.\ref{fig_vmdlinear} shows lower mobility (for linear polymers) than Fig.\ref{fig_vmdnetwork}
	(for cross-linked polymers).
		\begin{figure}[ht]
		\centering
		\includegraphics[width=0.9\linewidth]{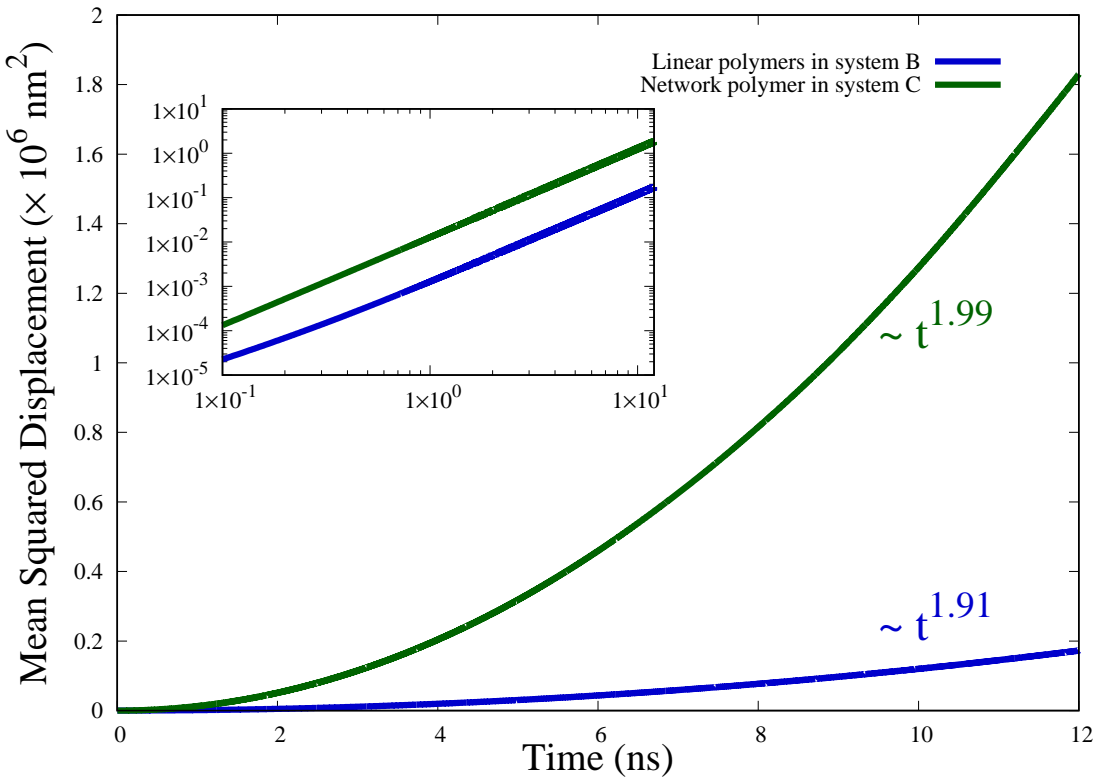}
		\caption{Mean Squared Displacement of polymers in systems B and C in absence of electric field. Inner is the same result but in logarithmic scale, indicating on power-law behavior of MSD.}
		\label{fig_msdPEO}
	\end{figure}

		\begin{figure}[ht]
		\centering
		\begin{subfigure}{0.4\linewidth}
			\centering
			\includegraphics[width=\linewidth]{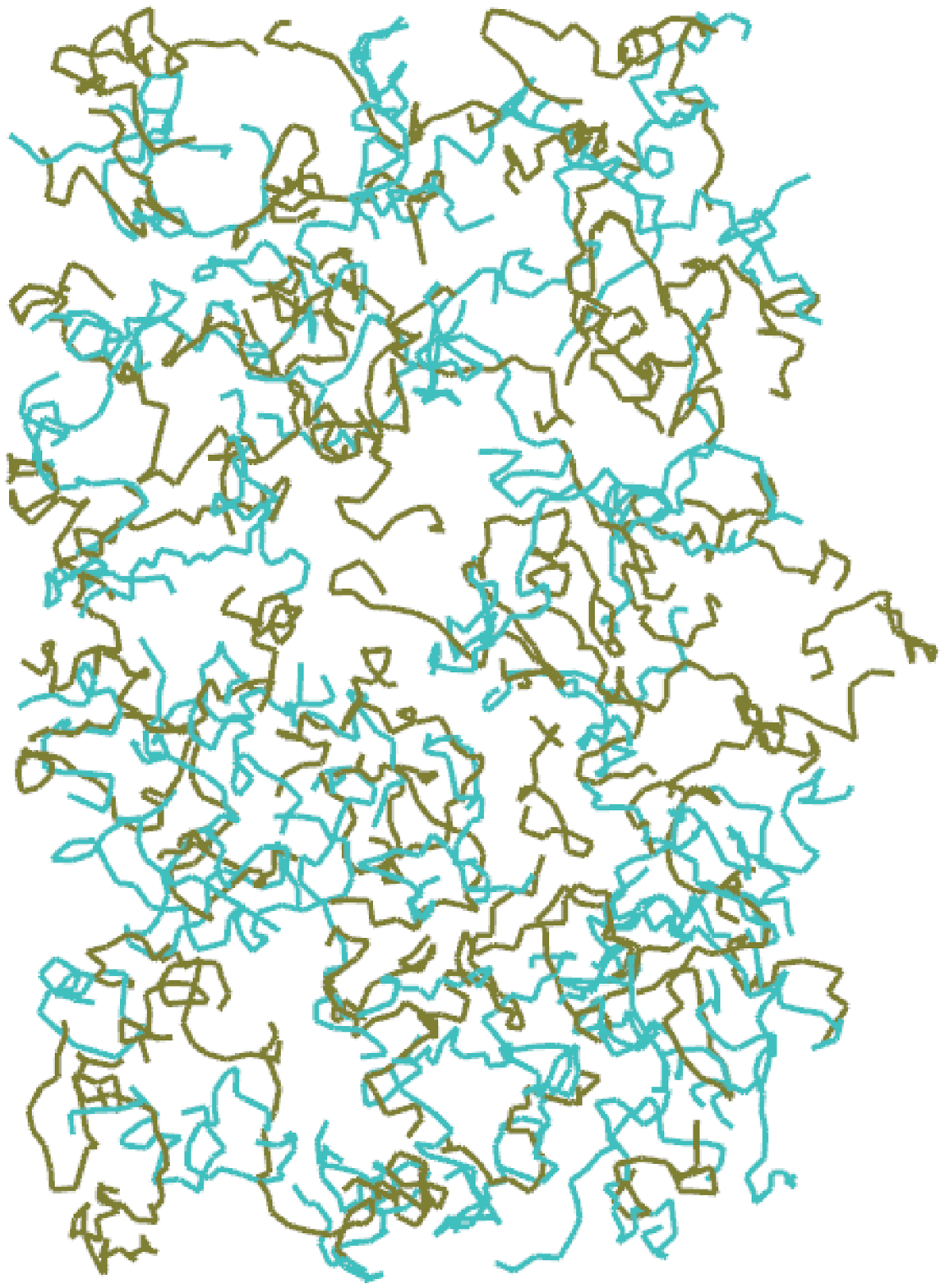}
			\caption{}
			\label{fig_vmdlinear}
		\end{subfigure}
		\hfill
		\begin{subfigure}{0.4\linewidth}
			\centering
			\includegraphics[width=\linewidth]{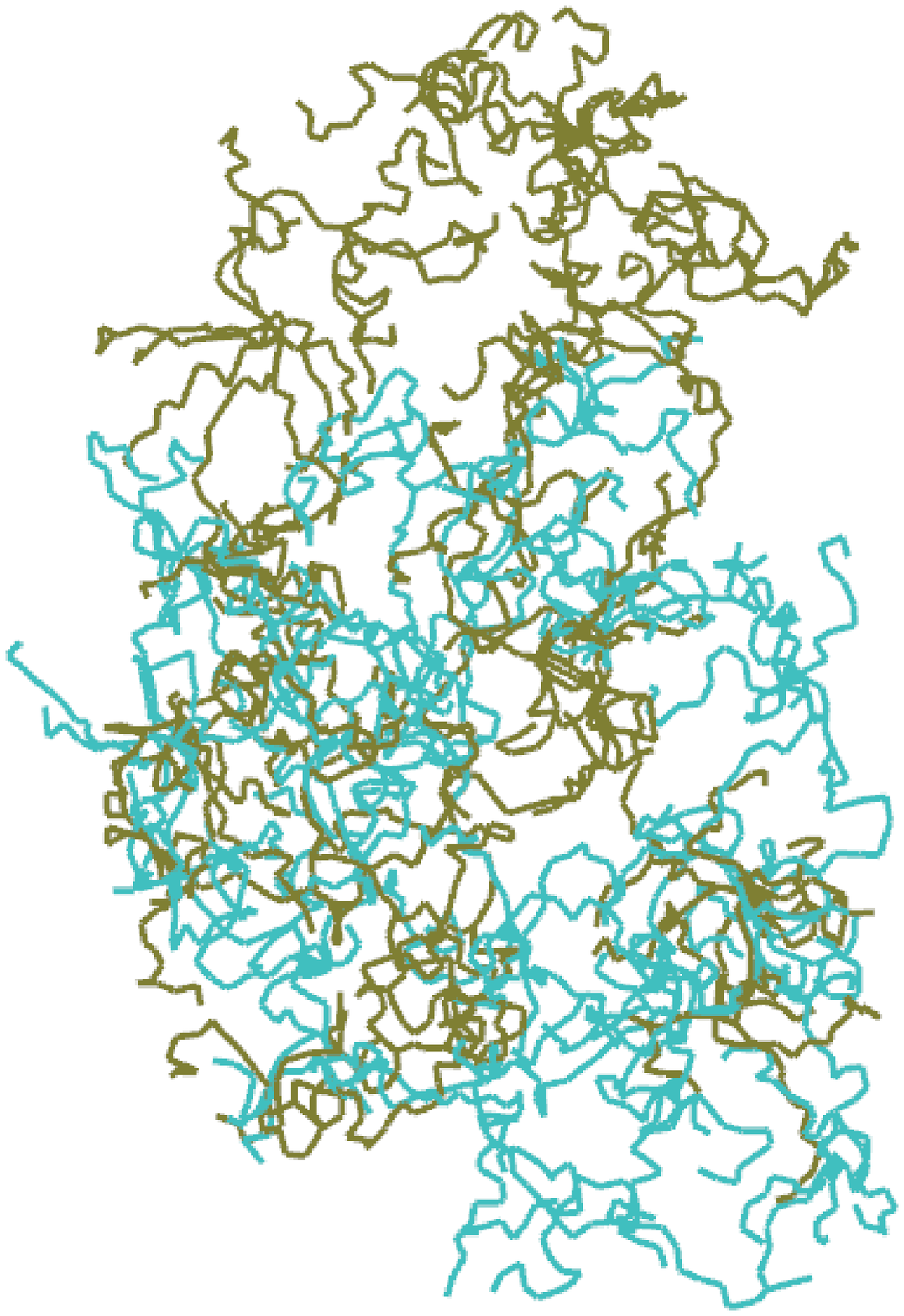}
			\caption{}
			\label{fig_vmdnetwork}
		\end{subfigure}
		\caption{Two different snapshots of polymers for (a) linear polymers in system B and (b) cross-linked polymers in system C.}
		\label{fig_vmd}
	\end{figure}
	
	In addition, we obtained MSD for cations and anions for all systems. Due to volume effects and the high mobility of cross-linked polymers, these polymers provide a limited path for charged particles in system C in comparison with linear polymers in system B. Therefore as shown in Fig.\ref{fig_msdIONS}, the charged particles' diffusion in system C is lower than in systems A and B. Since a lower diffusion coefficient is associated with a larger viscosity, the viscosity of system C is greater than systems A and B. Consequently, increasing the viscosity of the electrolyte improves its mechanical stability. Accordingly, system C has higher mechanical stability than systems A and B.
	
	Furthermore, it is important to note that cations and anions differ slightly in their results due to their mass differences. Compared to a cation, an anion has a lower weight, so its MSD plot shows a higher value.
		\begin{figure}[ht]
		\centering
		\includegraphics[width=0.9\linewidth]{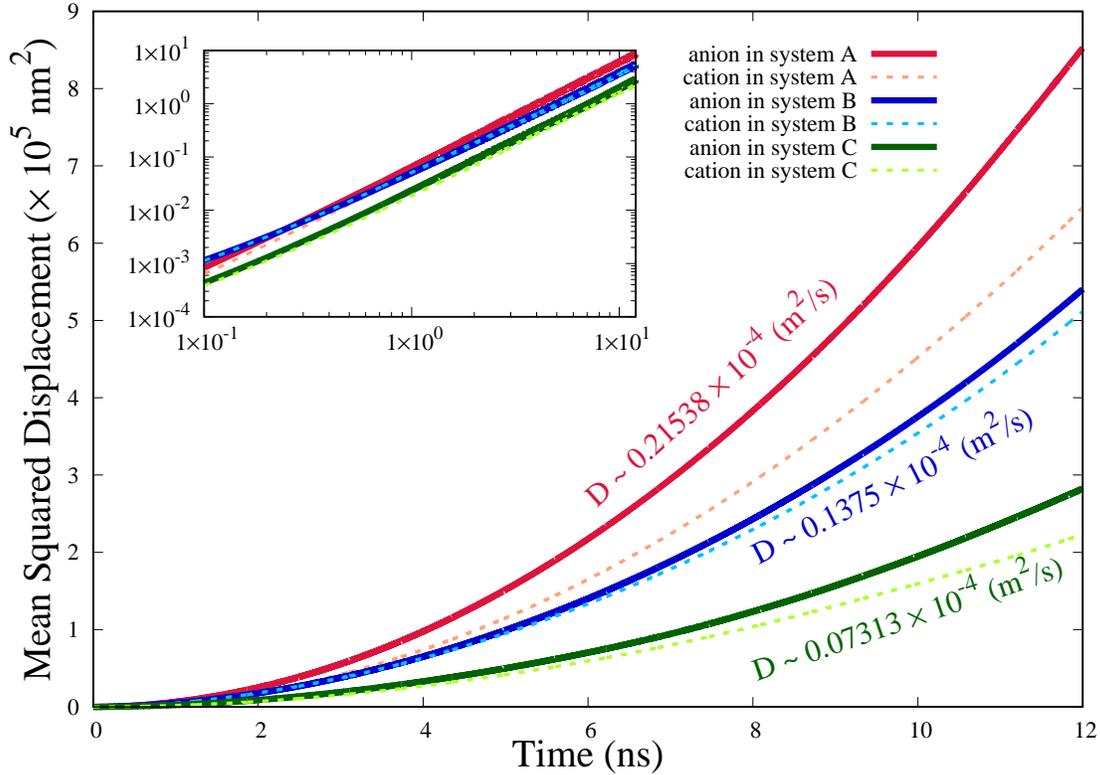}
		\caption{Mean Squared Displacement of ions in all three systems at $\Delta V=2~V$. Inner plot is the same as outer but
		in logarithmic scale.}
		\label{fig_msdIONS}
	\end{figure}
\section{Conclusion}
In summary, simulations demonstrate a difference between the performance of two types of electrolytes, i.e. liquid and solid electrolytes in supercapacitors. Solid electrolytes are used as an alternative to reduce the problems associated with liquid electrolytes. Therefore, simulation and comparison of these two categories can give us a clear insight into improving the efficiency of supercapacitors.

Liquid electrolyte-based supercapacitors have a smaller operating voltage window therefore, the amount of energy stored is less. Polymer electrolytes can be considered as a cluster of linear polymers or as cross-linked polymers in which linear polymers are connected and form a network. In linear polymers, since polymers can be present near the walls of the electrodes as well as inside the pores, the available space for ions in the vicinity of the electrodes is reduced. In contrast,  in a polymer network, the movement of polymers is collective, and therefore polymers can only be present in bulk space. Therefore, the accessible space for ions near the surfaces of the two electrodes and inside the pores increases. So the amount of charge stored in the electrodes and the supercapacitor efficiency increases.
\newpage
\bibliography{eyvazi.bib}
\end{document}